\documentclass{article}
\usepackage{graphicx}
\begin{document}

\title{Bounds for Entanglement of formation of two mode squeezed thermal States }
\author{\small  Xiao-Yu Chen\footnote{Email: drxychen@yahoo.com.cn; Address:129,College Road, Hangzhou, 310034,China } \\
\small Lab. of Quantum Communication and Computation, USTC,
230026, China;\\ \small \label{Correspnding address}School of
Science and Art,
China Institute of Metrology, Hangzhou, 310034,China; \\
 \small Peiliang Qiu \\
\small Department of Information Science and Electronic
Engineering,\\\small Zhejiang University, Hangzhou, 310027, China}
\date{}

\maketitle
\begin{abstract}
The upper and lower bounds of entanglement of formation are given
for two mode squeezed thermal state. The bounds are compared with
other entanglement measure or bounds. The entanglement
distillation and the relative entropy of entanglement of
infinitive squeezed state are obtained at the postulation of
hashing inequality.

\end{abstract}


\\

Quantum entanglement plays an essential role in all branches of the emerging
field of quantum information\cite{BennettDiVincenzo}. Recently a great deal of
attention has been devoted to the quantum information processing with
canonical continuous variables and various protocols for quantum communication
and computation are developed based on continuous variables \cite{Braunstein}
\cite{Lloyd} \cite{Gottesman}. With continuous variable optical system these
protocols can be implemented. The efficiency of entanglement manipulation
protocols critically depends on the quality of the entanglement that one can
generate. For practical purpose \cite{Scheel} one needs to determine the
entanglement degradation in transmission of two mode squeezed vacuum state
light through absorbing fibers. It is therefore essential to be able to
quantify the amount of entanglement in systems with continuous variables.

Of all the continuous variable quantum states, quantum Gaussian states are of
great practical importance, since these comprise essentially all the
experimentally realizable continuous variable states, theoretically they might
also be simple enough to be analytically tractable for the problems involved.
Gaussian state is completely specified by its mean $\eta$ and its correlation
matrix $\alpha$, where the mean can be dropped by local unitary operation so
that is irrelevant for entanglement problems. The density operator of Gaussian
state can be characterized \cite{Holevo} by its quantum characteristic
function $\chi\left(  z\right)  =\exp[i\eta^{T}z-\frac{1}{2}z^{T}\alpha z]$ ,
and any Gaussian state of two modes can be transformed into what we called the
\textit{standard form}, using local unitary operation only \cite{Duan}
\cite{Simon}. The corresponding characteristic function has mean $\eta=0$ and
the correlation matrix has the simple form
\begin{equation}
\alpha=\left[
\begin{array}
[c]{llll}%
b_{1}, & 0, & c_{1}, & 0\\
0, & b_{1}, & 0, & c_{2}\\
c_{1}, & 0, & b_{2}, & 0\\
0, & c_{2}, & 0, & b_{2}%
\end{array}
\right]  ,
\end{equation}

A particular simple subset of two mode Gaussian states is the two
mode squeezed thermal states (TMST), with $b_{1}=b_{2}=b,$
$c_{1}=-c_{2}=c$. The density operator of \ TMST is
$\rho_{ST}=S_{2}^{+}\left(  r\right)  \rho
_{0}S_{2}\left(  r\right)  ,$ where $S_{2}\left(  r\right)  =\exp[-r(a_{1}%
^{+}a_{2}^{+}-a_{1}a_{2})]$ is the two mode squeezed operator with
squeezing parameter $r\geq0$, and $\rho_{0}$ is the product of two
thermal states A and B whose average photon numbers are
$N_{A}=N_{B}=N$. The relationship between the two expressions of
TMST is $b=(N+1/2)\cosh2r,$ $c=(N+1/2)\sinh2r$. One can use the
unit scale parameters $\lambda=\tanh r$ and $v=N/(N+1)$, then the
inseparability criterion has a very simple form of
$\lambda>v$\cite{Duan} \cite{Simon}.

Even for the two parameter TMST, however, until now the amount of
the entanglement in several main measures remains an open
question. In this paper, I will study the upper and lower bounds
for the entanglement of formation of TMST. First I will establish
the lower bound of entanglement of formation, then the upper
bound, then compare them with other entanglement bounds and
measure, at last draw the conclusions.

Now let us consider the lower bound of entanglement of formation.
It was proved that entanglement of formation is nonincreasing
under local operation and classical communication \cite{Bennett}.
This presents us a nature way to obtain the lower bound of the
entanglement of formation of a mixed state. By performing a
special kind of local operation and (or) classical communication
to the state, converting it\ to a state whose entanglement of
formation is well known, the entanglement of formation of the
resultant state will provide a lower bound to that of the original
state.

For TMST, the simplest way of conversion is to map it on a
two-qubit system by means of local operation. One of the operation
for such purpose is with the Hermitian \textit{spin one-half}
operators\cite{Mista} $S_{j}^{\mu},$ $\mu=A,B$ for A, B parts
respectively,
$S_{1}^{\mu}+iS_{2}^{\mu}=2\sum_{m=0}^{\infty}\left\vert
2m\right\rangle _{\mu \mu}\left\langle 2m+1\right\vert
,S_{3}^{\mu}=\sum_{m=0}^{\infty }\left(  -1\right)  ^{m}\left\vert
m\right\rangle _{\mu \mu }\left\langle m\right\vert ,$which
satisfy the Pauli matrix algebra: $\left[
S_{i}^{\mu},S_{j}^{\nu}\right]  =2i\varepsilon_{ijk}%
\delta_{\mu\nu}S_{k}^{\mu}$.

Let us denote the mapped state with $\rho'$ and assign the
following qubit density matrix $\rho'_{A} $ to the original
reduced density matrix $\rho_{A}$, $\rho'_{A} =\frac{1}{2}\left(
I_{A}+\mathbf{S}^{A}\cdot\mathbf{\sigma}\right)  ,$where
$\mathbf{S}^{A}\mathbf{\cdot\sigma}=\sum_{i=1}^{3}Tr\left(  \rho_{A}S_{i}%
^{A}\right)  \sigma_{i}$, $\ \sigma_{i}$ are Pauli matrices and $I_{A}$ is the
identity operator on the Hilbert space of qubit A.

The map of bipartite density matrix is\cite{Mista} $$\rho'
=\frac{1}{4}\left(  I_{A}\otimes
I_{B}+\mathbf{S}^{A}\cdot\mathbf{\sigma
}\otimes I_{B}+I_{A}\otimes\mathbf{S}^{B}\cdot\mathbf{\sigma+}\sum_{i,j}%
t_{ij}\sigma_{i}\otimes\sigma_{j}\right)  $$, with
$t_{ij}=Tr\left(  \rho S_{i}^{A}S_{j}^{B}\right)  $ .

For TMST density matrix $\rho_{ST}$, its reduced state $\rho_{A}$ $=Tr_{B}%
\rho_{ST}$ turns out to be a one mode thermal state of the form $\left(  1-v'%
\right)  \sum_{m=0}^{\infty}v'^{m}\left\vert m\right\rangle \left\langle m\right\vert $ with $v'%
=\frac{v+\lambda^{2}}{1+v\lambda^{2}}$ . It is clear that
$Tr\left(  \rho _{A}S_{1,2}^{A}\right)  =0$, and a direct
calculation shows that $Tr\left( \rho_{A}S_{3}^{A}\right)
=\frac{1-v'}{1+v'}$. Meanwhile the two mode squeezing operator
$S_{2}\left(  r\right)  $ will generate or annihilate the same
number of bosons in mode A and mode B, it will not change the
difference of the occupation numbers of the two modes, one readily
gets $t_{ij}=0$, for $i\neq j$. As for the diagonal elements
$t_{ii}$,
from the definition of $t_{ij}$ , it follows that $t_{11}=-t_{22}%
=2Tr\sum_{m,n=0}^{\infty}\rho_{ST}\left\vert 2m,2n\right\rangle
\left\langle 2m+1,2n+1\right\vert .$Making use of the coherent
state representation of $\rho_{ST}$ ,after some algebra, one
obtains $t_{33}=\left(  \frac
{1-v}{1+v}\right)  ^{2}$ and%
\begin{equation}
t_{11}=2C_{0}\sum_{m,n=0}^{\infty}\sum_{l=0}^{2\min\left(
m,n\right) }\sqrt{\frac{2n+1}{2m+1}}\left(\begin{array}{c}
  2m+1 \\
  l+1
\end{array}\right)
\left(\begin{array}{c}
  2n \\
  l
\end{array}\right)\tau^{2l+1}\omega^{2m+2n-2l},
\end{equation}
with $C_{0}=\frac{\left(  1-v\right)  ^{2}\left(  1-\lambda^{2}\right)
}{1-v^{2}\lambda^{2}}$, $\tau=\frac{\lambda\left(  1-v^{2}\right)  }%
{1-v^{2}\lambda^{2}}$ , $\omega=\frac{v\left(  1-\lambda^{2}\right)  }%
{1-v^{2}\lambda^{2}}$.

The qubits map of TMST is%

\begin{equation}
\rho' =\left[
\begin{array}
[c]{llll}%
\frac{1+v^{2}\lambda^{2}}{\left(  1+v\right)  ^{2}\left(  1+\lambda
^{2}\right)  }, & 0, & 0, & \frac{t_{11}}{2}\\
0, & \frac{v}{\left(  1+v\right)  ^{2}}, & 0, & 0\\
0, & 0, & \frac{v}{\left(  1+v\right)  ^{2}}, & 0\\
\frac{t_{11}}{2}, & 0, & 0, & \frac{v^{2}+\lambda^{2}}{\left(  1+v\right)
^{2}\left(  1+\lambda^{2}\right)  }%
\end{array}
\right]
\end{equation}
The concurrence \cite{Wootters} of $\rho'$ will be
$C=\max\{0,t_{11}-2v/\left(  1+v\right)  ^{2}\}$. Denote $x=\frac
{1}{2}\left(  1+\sqrt{1-C^{2}}\right)  $, the entanglement of
formation of $\rho'$ reads%

\begin{equation}
E\left(  \rho'\right)  =h\left(  x\right)  =-x\log
x-(1-x)\log(1-x).
\end{equation}
Throughout this paper the base of $\log$ is $2$. One then obtains
the lower bound of entanglement of formation : $E_{lf}\left(
\rho_{ST}\right) =E\left(  \rho'\right)  $.

We now turn to the upper bound for entanglement of formation of
quantum Gaussian state and specifically TMST. The entanglement of
formation is defined as follows\cite{Bennett}. Given a density
matrix $\rho$ of bipartite quantum system, consider all possible
pure state decompositions of $\rho$, that is, all ensembles of
states $\left| \psi_{i}\right\rangle $ with probabilities $p_{i}$,
such that $\rho=\sum_{i}p_{i}\left|  \psi_{i}\right\rangle
\left\langle \psi_{i}\right| $. For each pure state $\left|
\psi\right\rangle $, the entanglement $E\left(  \psi\right)  $ is
defined as the entropy of the reduced state of $\left|
\psi\right\rangle $. The entanglement of formation of the mixed
state $\rho$ is then defined as the average entanglement of the
pure states of the decomposition, minimized over all
decompositions of $\rho:$ $E\left( \rho\right)
=\min\sum_{i}p_{i}E\left(  \psi_{i}\right) $.

From the definition, it is known that any ensemble as a decomposition of
$\rho$ has an average entanglement no less than the entanglement of formation
of $\rho$, and so that is a candidate in providing an upper bound for the
entanglement of formation of $\rho$ . The aim of us is to construct a series
of decompositions, and within the series find out the decomposition of minimum
average entanglement. Because the series is just a subset of the set of all
possible decompositions, the minimum within the series is a local minimum. It
provides a relatively tight upper bound for global minimum, the entanglement of
formation of $\rho$.

Among all the decompositions, Gaussian state decomposition is mathematically
easy to deal with. Rather than decompose a Gaussian state to its pure Gaussian
state ensemble, let us generate mixed Gaussian state from a pure Gaussian
seed. In the viewpoint of quantum channel, the way of generating mixed
Gaussian state from a pure Gaussian state is just\ the way of transmitting the
pure Gaussian state through quantum Gaussian channel \cite{Chen}. So that the
problem of how to generate mixed Gaussian state is physical, at least in
principle. Suppose a pure Gaussian state is given with correlation matrix
$\alpha_{s}$ and zero mean. Applying local unitary displacement operation on
it yields another pure Gaussian state, with the same correlation matrix but a
nonzero mean value or displacement $\eta$. The two states have the same amount
of entanglement because local unitary operation does not change entanglement
of pure state share by two parts A and B. If the displacement $\eta$ has a
Gaussian probability distribution, then the mixture of all these Gaussian
states will be a Gaussian state\cite{Holevo}. Let $P$ be a Gaussian
probability distribution with zero mean (for convenience) and covariance
matrix $M$ for variables $\left(  \eta_{A}^{q},\eta_{A}^{p},\eta_{B}^{q}%
,\eta_{B}^{p}\right)  $ =$\eta^{T}$. The quantum characteristic function of
the mixed state $\rho_{P}$ is%

\begin{equation}
\int\exp(i\eta^{T}z-\frac{1}{2}z^{T}\alpha_{s}z)P\left(  d\eta\right)
=\exp[-\frac{1}{2}z^{T}\left(  M+\alpha_{s}\right)  z].
\end{equation}
The correlation matrix of $\rho_{P}$ is $\alpha_{P}=M+\alpha_{s}$.
Normalizable of probability measure requires that $M$ should be positive
semi-definite, $M=0$ is included to correspond to $\delta$ function type of
probability distribution. So that arbitrary Gaussian mixed state $\rho_{P}$
can be decomposed into an ensemble of pure Gaussian states as far as the
difference of the correlation matrices is positive semi-definite. That is%

\begin{equation}
M=\alpha_{P}-\alpha_{s}\geq0. \label{wave0}
\end{equation}
So if the separable pure Gaussian state correlation matrix $\alpha_{s}$ is
found under the restriction of Eq. (\ref{wave0}), state $\rho_{P}$ should be
separable. If it is impossible to find such an $\alpha_{s}$, then one has to
turn to find an entangled pure Gaussian state correlation matrix $\alpha_{s}$
to fulfill Eq.(\ref{wave0}). The minimum of the entanglement of those pure
entangled Gaussian states will give out the upper bound of entanglement of
formation of state $\rho_{P}$.

Consider the entangled TMST $\rho_{ST}$ with $\lambda>v$. Let%

\begin{equation}
\alpha_{s}=\frac{1}{2}\left[
\begin{array}
[c]{llll}%
\cosh2x, & 0, & \sinh2x, & 0\\
0, & \cosh2x, & 0, & -\sinh2x\\
\sinh2x, & 0, & \cosh2x, & 0\\
0 & -\sinh2x, & 0, & \cosh2x
\end{array}
\right]  ,
\end{equation}
be exactly the squeezed vacuum state with squeezing parameter $x$.
$M\geq0$ requires that $b-\frac{1}{2}\cosh2x\geq0,(b-\frac{1}%
{2}\cosh2x)^{2}-(c-\frac{1}{2}\sinh2x)^{2}\geq0.$This can be
written as $\cosh2\left(  r-x\right)
\leq\frac{1+v^{2}}{1-v^{2}}.$Denote $v=\tanh
w$, one gets $r-x\leq w,$ or $x\geq r-w$. The minimum is $x_{0}%
=r-w$. Let $y=\sinh^{2}x_{0},$ in explicit form of $\lambda,v$, one
has $y=\frac{\left(  \lambda-v\right)  ^{2}}{\left(
1-\lambda^{2}\right)  \left(
1-v^{2}\right)  }$.\ The upper bound of entanglement of formation then will be%

\begin{equation}
E_{uf}\left(  \rho_{ST}\right)  =g\left(  y\right)  =\left(  y+1\right)
\log\left(  y+1\right)  -y\log y
\end{equation}

Let us compare the entanglement bounds obtained above with other bounds and
entanglement measure. First of them is the upper bound of relative entropy of
entanglement. The entanglement measure is defined by the minimization of the
distance of the entangled Gaussian state to the set\ of separable states
measured by the relative entropy, its bound was obtained by restricting the
minimization set to that of separable Gaussian states\cite{Scheel}. We can
further restrict \ it to the set of separable TMST $\widetilde{\rho}%
_{ST}=S_{2}\left(  \widetilde{r}\right)  \rho_{0}\left(  \widetilde{v}\right)
S_{2}^{+}\left(  \widetilde{r}\right)  $, with $\widetilde{\lambda}%
=\tanh\widetilde{r}\leq\widetilde{v}$. Then the minimization can
be
carried out. The upper bound will be%
\begin{equation}
E_{ur}\left(  \rho_{ST}\right) =-2g\left(  N\right) -2\log\left(
1-\widetilde{v}^{\ast}\right) -[\frac{1+v}{1-v}\cosh\left(
r-\widetilde{w}^{\ast}\right) -1]\log \widetilde{v}^{\ast}.
\end{equation}

The minimum reaches at $\widetilde{\lambda}=\widetilde{v}=\widetilde{v}^{\ast
}=\tanh\widetilde{w}^{\ast}$ for some $\widetilde{v}^{\ast}$.

The second one we should compare with is entanglement measure of logarithmic
negativity \cite{Vidal}, for TMST, it reads\qquad\qquad\qquad\qquad\qquad%
\begin{equation}
E_{LN}(\rho_{ST})=\log[(1+\lambda)(1-v)]-\log[(1-\lambda)(1+v)].
\end{equation}

The last one used in comparing is coherent information. The
coherent information of bipartite state $\sigma$ with reductions
$\sigma_{A}$ and $\sigma_{B}$ is defined as
\cite{Schumacher}$I^{\mu}\left(  \sigma\right) =S\left(
\sigma_{\mu}\right)  -S\left(  \sigma\right)  ,$for $S\left(
\sigma_{\mu}\right)  -S\left(  \sigma\right)  \geq0$ and
$I^{\mu}\left( \sigma\right)  =0$ otherwise, where $S$ denotes the
von Neumann entropy. And if for any bipartite state the one-way
distillable entanglement is no less than coherent information
which is called hashing inequality, then one obtains Shannon-like
formulas for the capacities \cite{Horodecki}. The hypothetical
hashing inequality reads%

\begin{equation}
D_{\rightarrow}\left(  \sigma\right)  \geq I^{B}\left(
\sigma\right)
\end{equation}
where $D_{\rightarrow}\left(  \sigma\right)  $ is forward
classical communication aided distillable entanglement of the
state. I do not mean to prove this inequality, but rather to check
if it is violated for some cases in continuous variable system. It
is known that two way distillation of entanglement is no less than
that of one way\cite{Bennett}, and entanglement of formation is no
less than relative entropy of entanglement\cite{Vedral1}, the
later is no less than distillation of entanglement\cite{Vedral2}.
In short one has $E_{f}\left(  \sigma\right)  \geq E_{r}\left(
\sigma\right) \geq D_{\leftrightarrow}\left(  \sigma\right) \geq
D_{\rightarrow
}\left(  \sigma\right)  $. Combining with hashing inequality\ one then obtains%

\begin{equation}
E_{uf}\left(  \sigma\right)  \geq E_{f}\left(  \sigma\right) \geq
I^{B}\left(  \sigma\right)  ,\quad E_{ur}\left( \sigma\right) \geq
E_{r}\left(  \sigma\right)  \geq I^{B}\left( \sigma\right)  .
\end{equation}
For TMST $\rho_{ST}$, \ the entropy of its reduced state is
$g\left(  N'\right)  $, with $N'
=\frac{v'}{1-v'}=\frac{v+\lambda^{2}}{\left(  1-v\right)  \left(
1-\lambda^{2}\right)  }$.
The coherent information is%

\begin{equation}
I^{B}\left(  \rho_{ST}\right)  =\max\{g\left(  N'%
\right)  -2g\left(  N\right)  ,0\}
\end{equation}

By comparison of all bounds and measure of TMST I find that: (1) the upper
bound of entanglement of formation and of relative entropy of entanglement are
no less than coherent information, so it can be anticipated that hashing
inequality is also true for TMST or more generally for bipartite Gaussian
state; (2) As the squeezing parameter tends towards zero, the lower and upper
bounds of entanglement of formation tend to coincide with each other. (3) As
the squeezing parameter tends towards infinitive, the upper bound of relative
entropy of entanglement tends to coincide with the coherent information, so if
the hashing inequality is right, the relative entropy of entanglement, the
distillation entanglement and coherent information are all equal for
infinitive squeezing state. (4) The entanglement measure of logarithmic
negativity is at the topmost of all the bounds.

The main points of this paper are as follow: (1) The lower bound
of entanglement of formation of TMST is drawn from the fact that
local operation can only decreases if not preserves the
entanglement, at lower squeezing side, it almost coincides with
the upper bound. This result leads to the possible determination
of the entanglement of formation at lower squeezing level. (2)As
for the upper bound of entanglement of formation of quantum
Gaussian state, an inequality is derived, for the case of TMST, an
analytical expression is given out. (3)By carrying out the
minimization of the upper bound of relative entropy of
entanglement\cite{Scheel} for TMST I determine the entanglement
distillation and the relative entropy of entanglement of TMST with
infinitive squeezing at the postulation of hashing inequality.

\begin{figure}
[ptb]
\begin{center}
\includegraphics[
trim=0.000000in 0.000000in -0.157259in 0.000000in,
height=2.0081in, width=2.5097in
]%
{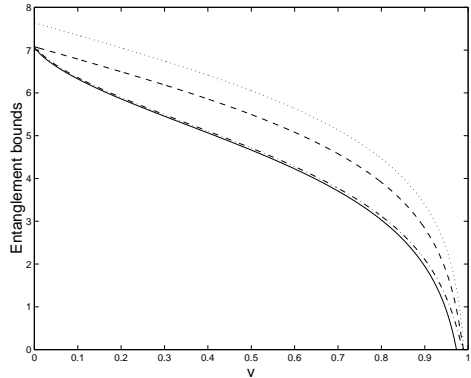}%
\caption{Comparison of bounds and measure at $\lambda\rightarrow1$
side. $\lambda=0.99,$ dashed line for $E_{uf}$, dashdot for
$E_{ur}$, solid for
$I^{B}$, dotted for $E_{LN}$}%
\end{center}
\end{figure}

\begin{figure}
[ptb]
\begin{center}
\includegraphics[
trim=0.000000in 0.000000in 0.011666in 0.000000in, height=2.0081in,
width=2.5097in
]%
{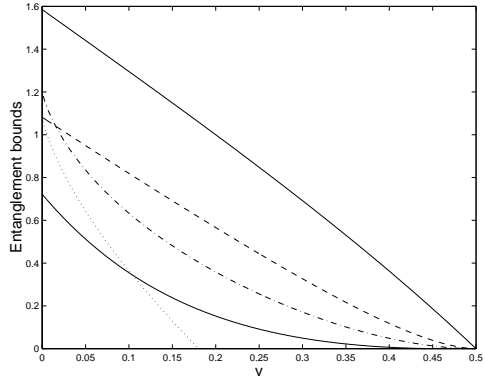}%
\caption{Comparison of bounds and measure at $\lambda=0.5,$ dashed
line for $E_{uf}$, dashdot for $E_{ur}$, the lower\ solid for
$E_{lf}$, dotted for
$I^{B}$. the topmost solid for $E_{LN}$.}%
\end{center}
\end{figure}

\begin{figure}
[ptb]
\begin{center}
\includegraphics[
trim=0.000000in 0.000000in -0.138042in 0.000000in,
height=2.0081in, width=2.5097in
]%
{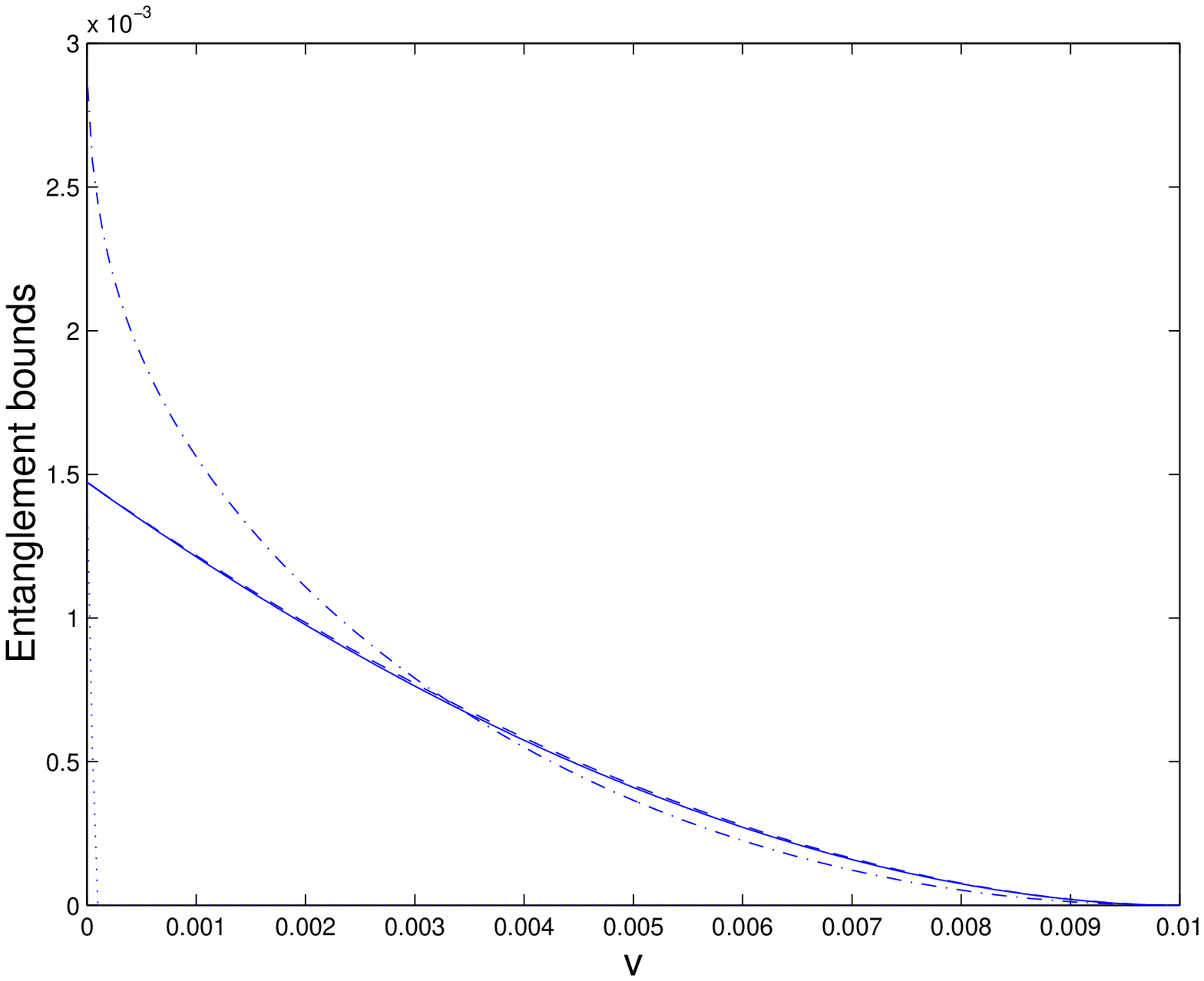}%
\caption{Comparison of bounds at $\lambda\rightarrow0$ side.
$\lambda=0.01,$ dashed line for $E_{uf}$, dashdot for $E_{ur}$,
solid for $E_{lf}$, dotted for$I^{B}$. Entanglement measure
$E_{LN}$ is omitted for significantly greater
than the bounds.}%
\end{center}
\end{figure}


\begin{thebibliography}{99}                                                                                               %


\bibitem {BennettDiVincenzo}C. H. Bennett, Phys. Today \textbf{48}, No 10, 24
(1995); D. P. DiVincenzo, Science \textbf{270}, 255 (1995).

\bibitem {Braunstein}S. L. Braunstein and H. J. Kimble, Phys. Rev. Lett.
\textbf{80}, 869 (1998); S. L. Braunstein, Nature. \textbf{394}, 47 (1998).

\bibitem {Lloyd}S. Lloyd and S. L. Braunstein, Phys. Rev. Lett. \textbf{82},
1784 (1999).

\bibitem {Gottesman}D. Gottesman, A. Kitaev and J. Preskill, Phys. Rev. A
\textbf{64}, 012310 (2001).

\bibitem {Scheel}S. Scheel and D.-G. Welsch, arXiv: quant-ph/0103167.

\bibitem {Holevo}A. S. Holevo, M. Sohma and O. Hirota, Phys. Rev. A
\textbf{59}, 1820 (1999).

\bibitem {Duan}L. M. Duan, G. Giedke, J. I. Cirac and P. Zoller, Phys. Rev.
Lett. \textbf{84}, 2722 (2000).

\bibitem {Simon}R. Simon, Phys. Rev. Lett. \textbf{84}, 2726 (2000)

\bibitem {Bennett}C. H. Bennett, D. P. DiVincenzo, J. A. Smolin and W. K.
Wootters, Phys. Rev. A \textbf{54}, 3824 (1996).

\bibitem {Mista}L. Mi\v{s}ta, Jr. R. Filip and J. Fiur\'{a}\v{s}ek, Phys. Rev.
A \textbf{65}, 062315 (2002).

\bibitem {Wootters}W. K. Wootters, Phys. Rev. Lett. \textbf{80}, 2245 (1998).

\bibitem {Chen}X.Y. Chen and P. L. Qiu, Chin. Phys. \textbf{10}, \ 779 (2001)

\bibitem {Vidal}G. Vidal and R. F. Werner, Phys. Rev. A \textbf{65}, 032314 (2002).

\bibitem {Schumacher}B. Schumacher, Phys. Rev. A \textbf{54}, 2614 (1996); B.
Schumacher and M. A. Nielsen, Phys. Rev. A \textbf{54}, 2629 (1996).

\bibitem {Horodecki}M. Horodecki, P.Horodecki and R.Horodecki, Phys. Rev.
Lett. \textbf{85}, 433 (2000).

\bibitem {Vedral1}V. Vedral and M. B. Plenio, Phys. Rev. A \textbf{57}, 1619 (1998).

\bibitem {Vedral2}V. Vedral, Rev. Mod. phys. \textbf{74}, 197 (2002).

\end{thebibliography}
\end{document}